\documentstyle[11pt,psfig,epsf]{article}

\def\be{\begin{eqnarray}}
\def\ee{\end{eqnarray}}
\def\ben{\begin{enumerate}}\def\een{\end{enumerate}}
\def\der{\mbox{d}}
\def\d {\partial}
\def\L{{\cal L}}
\def\tr{{\rm tr}}
\def\prl {Phys. Rev. Lett.}\def\pr{Phys. Rev.}
\def\np{Nucl. Phys.}\def\pl{Phys. Lett.}
\def\la{\langle}\def\ra{\rangle}
\def\O{{\cal O}}
\def\roughly#1{\mathrel{\raise.3ex\hbox{$#1$\kern-.75em%
\lower1ex\hbox{$\sim$}}}}

\def\lsim{\roughly<}
\def\gsim{\roughly>}
\def\f{{f_\pi^\star}}
\def\del{\partial}
\renewcommand{\thefootnote}{\fnsymbol{footnote}}
\begin{document}

\begin{titlepage}
\hfill{SNUTP 97-052}
\vspace{.3cm}
\begin{center}
\ \\
{\Large \bf FLUCTUATIONS IN\\  ``BR-SCALED'' CHIRAL LAGRANGIANS}

\ \\
\vspace{0.2cm}
{Chaejun Song$^{(a)}$\footnote{E-mail: chaejun@fire.snu.ac.kr}, 
G.E. Brown$^{(b)}$\footnote{E-mail: gbrown@insti.physics.sunysb.edu},
Dong-Pil Min$^{(a)}$\footnote{E-mail: dpmin@phya.snu.ac.kr} and 
Mannque Rho$^{(c,d)}$\footnote{E-mail: rho@wasa.saclay.cea.fr}}

\vskip 0.5cm

{\it (a) Department of Physics and Center for Theoretical Physics,}

{\it Seoul National University, Seoul 151-742, Korea}

{\it (b) Department of Physics,  State University of New York,}

{\it Stony Brook, NY 11794, U.S.A.}

{\it (c) Departament de Fisica Te\`orica, Universitat de Val\`encia,}

{\it 46100 Burjassot (Val\`encia), Spain} 

{\it (d) Service de Physique Th\'eorique, CEA Saclay,}

{\it F-91191 Gif-sur-Yvette, France\footnote{Permanent address.}}
\end{center}
\vskip 0.5cm
\centerline{\bf ABSTRACT}
\vskip 0.6cm
We develop arguments for ``mapping"  the effective chiral Lagrangian whose
parameters are given by ``BR" scaling to a Landau Fermi-liquid 
fixed-point theory for nuclear matter in describing fluctuations
in various flavor (e.g., strangeness) directions. We use for this purpose the 
effective Lagrangian used by Furnstahl, Tang and Serot that incorporates the
trace anomaly of QCD in terms of a light-quark (quarkonium) 
degree of freedom with the
heavy (gluonium) degree of freedom integrated out. The large anomalous 
dimension $d_{an}\approx 5/3$ for the scalar field found by Furnstahl
et al to be needed for a correct description of nuclear matter 
is interpreted as an indication for a strong-coupling regime and the ground 
state given by the BR-scaled parameters is suggested as the background around 
which fluctuations can be rendered weak so that mean-field approximation
is reliable. We construct a simple model with BR scaled parameters that
provides a satisfactory description of the properties of matter at 
normal nuclear matter density. Given this,
fluctuations around the BR scaled background are dominated by tree diagrams.
Our reasoning relies heavily on recent developments in the study
of nucleon and kaon properties in normal and dense nuclear matter, e.g., 
nucleon and kaon flows in heavy-ion processes, kaonic atoms, 
and kaon condensation 
in dense compact star matter.
\vspace{2cm}
\end{titlepage}
\setcounter{footnote}{0}
\renewcommand{\thefootnote}{\arabic{footnote}}
\section{Introduction}
\indent\indent
In computing fluctuations in various flavor directions (such as strangeness)
in nuclear processes, the standard procedure has been to assume that the ground
state of hadronic matter is given by
the conventional nuclear matter and append in an arbitrary fashion flavor  
fluctuations on top of the assumed ground state using effective chiral 
Lagrangians at low chiral order. In doing this, 
one usually takes a theory for the ground state
from standard many-body treatment and adds mesonic fluctuations
using a chiral Lagrangian with however no constraints imposed for 
consistency between the ground state and the 
fluctuations. This is clearly an unsatisfactory procedure for going 
beyond the normal matter condition
 although with some astute intuitive input, 
one can make a fairly successful phenomenology of a variety of
meson-fluctuation processes at the normal matter density.

In this paper, we make a first step toward bridging the physics of the ground
state to that of fluctuations on top of it in the framework of an 
effective chiral Lagrangian field theory.

The problem can be 
stated as follows. Suppose one wants to describe the property 
of light-quark mesons in dense nuclear medium as, for instance,
probed in dilepton productions
in heavy-ion collisions (e.g., CERES) or in electroproduction of vector mesons
 inside nuclear medium (e.g., CEBAF). As has been recently shown by Li, Ko and
Brown \cite{LKB}, such a process can
be most economically -- and remarkably well --
described in terms of a chiral Lagrangian in mean field with the 
parameters of the Lagrangian scaled according to ``Brown-Rho (BR) scaling"
\cite{BR91}.  In this approach, however, one treats the ``matter" 
(i.e., nuclear) property in a way disconnected from -- 
although not inconsistent
with -- the BR-scaled chiral Lagrangian that is used to describe the vector 
meson property. The underlying assumption here is that the ground state is
given by the same {\it effective} chiral Lagrangian which is supposed to
include high-order quantum corrections, perhaps as 
a ``chiral liquid" as suggested by Lynn \cite{lynn} or as mean field
of the BR-scaled chiral Lagrangian as suggested in \cite{BR96} 
(``BR conjecture"). It is not yet fully understood how the Fermi surface is obtained in this scheme. However given the matter with a 
Fermi surface given by such a description, one can then {\it map} the
BR-scaled chiral Lagrangian to Landau Fermi liquid fixed point theory in the
way explained 
in \cite{FR96}. This mapping has been tested and found to be 
phenomenologically successful in such static
properties of nuclei as the nuclear gyromagnetic ratio, $g_A^\star$, the 
nucleon effective mass $m_N^\star$ etc\cite{FR96}. We take this success
as the first justification for the ``BR conjecture." This provides
a link between the baryon property and meson property inside dense medium.
It also enables one to extrapolate from normal nuclear matter at 
equilibrium to hadronic matter under extreme conditions.

A further support comes from processes involving kaons in nuclear matter.
Given the ground state of the matter with the scaled parameters, fluctuations
on top of it into the kaonic flavor direction seem to 
give correct properties of the 
$K^\pm$ in medium as seen in kaonic atom, subthreshold productions
and flows of $K^\pm$ in heavy-ion collisions (e.g., KaoS and FOPI) 
\cite{gebgsi}.
We take this as the second justification.

A basic problem however remains
when we apply the theory to kaon condensation in compact-star matter, one
of the most fascinating phenomena associated with strangeness in dense
matter. Here one is
dealing with a change of the ground state from that of nonstrange to
strange matter and hence the whole system, that is, the bulk involving  
the ground state and excitations on top of it,
has to be treated on the same footing.  In works up to date \cite{KN,LBMR},
this matter has not been consistently treated.
It is the aim of this paper to attempt to remedy this defect.

This paper is organized as follows. In Section 2, a general strategy of an
effective chiral Lagrangian as applied to dense medium is presented and the
model of Furnstahl et al \cite{tang} (referred to as FTS1)
that incorporates both chiral symmetry
and the trace anomaly of QCD is presented in this framework. In Section 3, the
role of anomalous dimension for the scalar field that enters into the 
trace anomaly of the FTS1 model on the structure of many-body forces and
the compression modulus of nuclear matter is examined. Section 4 is
devoted to the proposition that the mean field theory with the FTS1 Lagrangian
corresponds to Lynn's nontopological soliton  or a chiral
liquid. We discuss how this chiral liquid can be identified with Landau's
Fermi liquid structure of drop of nuclear matter in terms of renormalization
group flow arguments using developments in condensed matter physics. 
In Section 5,
BR scaling is incorporated into a chiral Lagrangian to obtain a weak-coupling
description of the same physics as the (strong-coupling)
FTS1 mean-field theory. This defines the background
at a given finite density around which fluctuations can be made.
In Section 5, the BR-scaled parameters introduced in the previous section
can be mapped to Landau Fermi liquid parameters and a contact with low-energy 
nuclear properties as well as kaon-nuclear interactions at normal matter
and higher densities be made through the mapping of the parameters.
A summary and conclusions are given in Section 6. The appendix shows
how sensitive the EOS is to the correlation parameters for $\rho > \rho_0$.

\section{Effective Chiral Lagrangian for Nuclear Matter}
\label{strategy}
\indent\indent
We begin  by recalling the main result of
 \cite{BR96}. Let an effective Lagrangian ${{\L}^{eff}}$
be defined as
\be
S^{eff} =\int d^4x\L^{eff}
\ee
where $S^{eff}$ is
a Wilsonian effective action arrived at after integrating out 
high-frequency modes of the nucleon and other heavy degrees of freedom.
This action is then given in terms of sum of terms organized in chiral order
in the sense of effective theories at low energy. The key point of
Ref. \cite{BR96} is that the mean-field solution of the chiral effective
Lagrangian with the parameters given by the BR scaling \cite{BR91} approximates the solution
\be
\delta S^{eff}=0.\label{extreme}
\ee

Our ultimate aim in this paper (and subsequent papers)
is to ``derive" the results of 
Refs. \cite{BR96,gebgsi}
starting with a chiral Lagrangian description of the ground state as
specified above around which
fluctuations in various flavor sectors are to be made.
To do this, we take a phenomenologically successful mean-field model
of Walecka type to describe the ground state.
In a recent publication, Furnstahl, Tang and Serot  \cite{tang} 
constructed an effective {\it quantum} nonlinear chiral model that in mean field
reproduces quite well all basic nuclear properties. This model that we shall
refer to as FTS1 model incorporates the trace anomaly of QCD in terms of a light
(``quarkonium") scalar field $S$ and a heavy (``gluonium") scalar field $\chi$. In a general framework of chiral dynamics, it is possible to
avoid the use of the conformal anomaly of QCD by appealing to
other notions of effective field theories such as ``naturalness condition"
as in \cite{tang2} (that we shall refer to as FTS2) leading to 
an effective mean field theory which gives an equally satisfactory
phenomenology as the FTS1. For our purpose, however, 
it proves to be more convenient to exploit the role of the light
scalar field that figures in the trace anomaly. In particular,
it makes the successful
description of the nucleon flow in heavy-ion collisions obtained by 
Li et al \cite{LBLK96} (who use the FTS1 theory) more readily understandable.

As in FTS1, we shall assume the heavy scalar 
field to have the canonical scale dimension
($d=1$) while the light scalar field is taken to transform under scale
transformation as 
\be
S(\lambda^{-1}x)=\lambda^d S(x)
\ee
with $d$ a {\it parameter} that can differ from unity, the canonical
dimension. The assumption here is that radiative corrections in the scalar 
channel can be summarized by an anomalous dimension $d_{an}=d-1\neq 0$. 
A heuristic justification for this assumption will be given below 
in terms of a renormalization group flow argument.
One further assumption that FTS1 adopt from  Ref. \cite{miransky} is
that there is 
no mixing between the light scalar $S(x)$ and the heavy scalar $\chi$
in the trace anomaly. Their Lagrangian has the form
\be
\L^{eff}=\L_s-H_g\frac{\chi^4}{\chi_0^4}(\ln\frac{\chi}{\chi_0}-\frac{1}{4})
-H_q(\frac{S^2}{S_0^2})^{\frac{2}{d}}(\frac{1}{2d}\ln\frac{S^2}{S_0^2}
-\frac{1}{4})
\ee
where $\L_s$ is the chiral- and scale-invariant Lagrangian containing
$\chi ,S,N,\pi , \omega$, etc. Here $\chi_0$ and $S_0$ are the vacuum
expectation values with the vacuum $|0\ra$ defined in the matter-free space:
\be
\chi_0\equiv\la 0|\chi|0\ra, \ \ \ \ S_0\equiv\la 0|S|0\ra.
\ee
The trace of the improved energy-momentum 
tensor \cite{coleman} from the Lagrangian is;
\be
\d_{\mu}D^{\mu}=\theta_\mu^\mu = -H_g\frac{\chi^4}{\chi_0^4}
-H_q(\frac{S^2}{S_0^2})^{2/d}
\ee
where $D_{\mu}$ is the dilatation current. The mass scale associated
with the gluonium degree of freedom is higher than that of chiral symmetry,
$\Lambda_\chi\sim 1$ GeV, so it is integrated out in favor of the light
scalar in which case the FTS1 effective Lagrangian takes the form 
\be
\L&=&\bar{N}[i\gamma_{\mu}(\d^{\mu}+iv^{\mu}
+ig_v\omega^{\mu}+g_A\gamma_5 a^{\mu})
-M+g_s\phi )N
\nonumber\\
& &-\frac{1}{4}F_{\mu\nu}F^{\mu\nu}+
\frac{1}{4!}\zeta g_v^4 (\omega_{\mu}\omega^{\mu})^2 
\nonumber\\
& &+\frac{1}{2}(1+\eta 
\frac{\phi}{S_0})[\frac{f_{\pi}^2}{2}\tr(\d_{\mu}U\d^{\mu}U^{\dagger})+
m_v^2 \omega_{\mu}\omega^{\mu}]\nonumber\\
& &+\frac{1}{2}\d_{\mu}\phi\d^{\mu}\phi
-\frac{m_s^2}{4}S_0^2d^2\{(1-\frac{\phi}{S_0})^{4/d}[\frac{1}{d}
\ln(1-\frac{\phi}{S_0})-\frac{1}{4}]+\frac{1}{4}\}\label{leff}
\ee
where $S=S_0-\phi$, $\eta$ and $\zeta$ are unknown parameters 
to be fixed and
\be
\xi^2 &=& U =e^{i\vec{\pi}\cdot\vec{\tau}/f_\pi}\nonumber\\
v_\mu &=& -\frac{i}{2}(\xi^{\dagger}\d_\mu\xi 
+\xi\d_\mu\xi^\dagger )\nonumber\\
a_\mu &=& -\frac{i}{2}(\xi^\dagger\d_\mu\xi
-\xi\d_\mu\xi^\dagger ).\nonumber
\ee

It is important to note that
the FTS1 Lagrangian is an effective (quantal) Lagrangian in the sense
specified above. The effect of 
high-frequency modes of the nucleon field and other massive degrees of freedom
appears in the parameters of the Lagrangian and in the counter terms that
render the expansion meaningful. It presumably includes also vacuum 
fluctuations in the Dirac sea of the nucleons \cite{tang,tang3}. In general,
it must be a lot more complicated. Indeed,
 one does not yet know how to implement this strategy in 
full rigor given that one does not know what the matching conditions are.
In \cite{tang,tang2}, {\it the major work is, however, done by choosing 
to fit the
relevant parameters of the FTS1 Lagrangian to empirical informations}.

The energy density for uniform nuclear matter with the static mean fields
obtained from (\ref{leff}) is 
\be
\varepsilon
 &=& \frac{\gamma}{(2\pi)^3}\int^{k_F}d^3k \sqrt{\vec{k}^2+(M-g_s\phi_0)^2}
\nonumber\\
& & -\frac{m_v^2}{2}(1+\eta\frac{\phi_0}{S_0}) \omega_0^2
+g_v\rho_B \omega_0-\frac{\zeta}{4!}g_v^4\omega_0^4\nonumber\\
& &+\frac{m_s^2}{4}S_0^2d^2\{(1-\frac{\phi_0}{S_0})^{4/d}
[\frac{1}{d}\ln(1-\frac{\phi_0}{S_0})-\frac{1}{4}]+\frac{1}{4}\}.
\label{energydensity}
\ee
Here $\gamma$ is the degeneracy factor.
\section{Anomalous Dimension}
\indent\indent
The best fit to the properties of nuclear matter and finite nuclei is 
obtained with the parameter set T1 when the scale 
dimension of the scalar $S$ is near $d=2.7$. In this section, we analyze
how this comes out and present what we understand of the role of
the large anomalous dimension $d_{an}=d-1\approx 1.7$ in nuclear dynamics. 
In what follows, the 
parameter T1 with this anomalous dimension will be taken as a canonical
parameter set\footnote{Explicitly the T1 parameters are: $d=2.7$, $g_s^2=99.3$, $m_S=509$ MeV, $S_0=90.6$ MeV, $g_V^2=154.5$, $\xi=0.0402$
and $\eta=-0.496$.}.

\subsection{Scale anomaly}
\indent\indent
Following Coleman and Jackiw \cite{coleman}, the scale anomaly  can be discussed
in terms of an anomalous Ward identity.  Define 
$\Gamma_{\mu\nu}(p,q)$ and $\Gamma(p,q)$ by 
\be
G(p)\Gamma_{\mu\nu}(p,q)G(p+q)
&=&\int d^4xd^4ye^{iq\cdot x}e^{ip\cdot y}
\langle 0\mid T^\ast \theta_{\mu\nu}(x)\varphi (y)\varphi (0)
\mid 0\rangle\\
G(p)\Gamma_G(p+q)
&=&\int d^4xd^4ye^{iq\cdot x}e^{ip\cdot y}
\langle 0\mid T^\ast\d_\mu D^{\mu}\varphi (y)\varphi (0)
\mid 0\rangle
\ee
with the renormalized propagator $G(p)$ and the renormalized 
fields $\varphi (x)$. Here $T^\ast$ is the covariant T-product and
$D_{\mu}(x)$ the dilatation current. 
A naive consideration on Ward identities would give 
\be
g_{\mu\nu}\Gamma^{\mu\nu}(p,q)
=\Gamma (p,q)-idG^{-1}(p)-idG^{-1}(p+q)
\ee
with $d$ the scale dimension of $\varphi (x)$. However $\Gamma$ is ill-defined
due to singularity and so has to be regularized. With the regularization,
the Ward identity reads
\be
g_{\mu\nu}\Gamma^{\mu\nu}(p,q)
&=&\Gamma (p,q)-idG^{-1}(p)-idG^{-1}(p+q)+A(p,q)\\
A(p,q)&\equiv &\lim_{\Lambda\rightarrow\infty}
\Gamma (p,q,\Lambda )-\Gamma (p,q)
\ee
where the additional term, $A$, is the anomaly. This anomaly corresponds
to a shift in the dimension of the field involved at the lowest loop
order but at higher orders there are vertex corrections. One obtains
however a simple result when the beta functions vanish  
at zero momentum transfer\cite{coleman}.
Indeed in this case, the only effect of the anomaly will appear as an
anomalous dimension.  In general this simplification does not occur.
However it can take place when there are nontrivial fixed points in the theory.
Now using the reasoning developed in condensed matter physics \cite{shankar},
we argue as in \cite{FR96} and elaborated further later
that nuclear matter is given in the absence of BCS channel 
by a Landau Fermi liquid fixed point theory 
with vanishing beta functions of the four-Fermi interactions
and that all quantum fluctuation effects would appear in the anomalous
dimension of the scalar field $S$. That nuclear matter is a Fermi liquid
fixed point seems to be well verified at least 
phenomenologically as first suggested in \cite{FR96}.
However that fluctuations into the scalar channel can be summarized into
an anomalous dimension is a conjecture that requires a proof.
We conjecture here that this is one way we can understand the success of 
the FTS1 model.

\subsection{Nuclear matter properties at $d_{an}\approx 5/3$}
\indent\indent

The FTS1 theory has some remarkable features  
associated with the large anomalous dimension. Particularly striking
is the dependence on the anomalous dimension of the compression modulus
and many-body forces.
\subsubsection{Compression modulus $K$}
\begin{figure}[tbh]
\setlength{\epsfysize}{3in}
\centerline{\epsffile{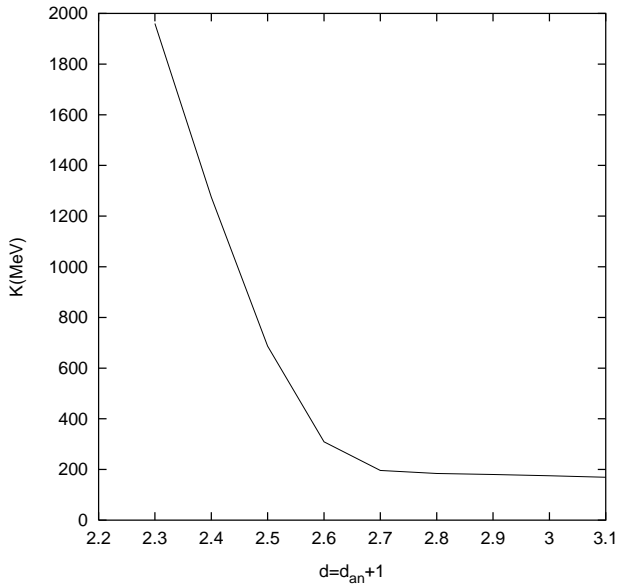}}
\caption[]{Compression modulus vs. anomalous dimension. The 
parameter set used here 
is the T1 in FTS1. This shows the sensitivity of the 
compression modulus to the anomalous dimension.}\label{K}
\end{figure}

\indent\indent
In Table \ref{p} are listed the compression modulus $K$ and the equilibrium
Fermi momentum $k_{eq}$ vs. the $d$ of the scalar field $\phi$.
As the $d$ increases, the $K$ drops very rapidly and stabilizes at $K\sim
200$ MeV for $d\approx 2.6$ and stays nearly constant for $d>2.6$.
This can be seen in Figure \ref{K}. The equilibrium Fermi momentum 
on the other hand slowly decreases uniformly as the $d$ increases.

Unfortunately, we have no simple understanding on the mechanism that makes
the compression modulus $K$ stabilize at the particular value $d_{an}\approx
5/3$. We believe there is a non-trivial correlation between
this behavior of $K$ and the observation made below
that the scalar logarithmic interaction brought in by the 
trace anomaly is entirely given {\it at the saturation point}
by the quadratic term at the same $d_{an}$ with the higher polynomial
terms (i.e., many-body interactions) contributing more repulsion for increasing
anomalous dimension. 
At present our understanding is purely numerical and hence incomplete.
The results of the extensive numerical analyses we have performed
and our interpretation thereof 
will be reported elsewhere \cite{SBMR2}.

\begin{table}[tbh]
\caption[]{Equilibrium Fermi momentum $k_{eq}$ and binding energy 
$B=M-E/A$ as a function of $d$ for Figure \ref{K}}\label{p}
\vskip .3cm
\begin{center}
\begin{tabular}{|c|c|c|c|}\hline
$d$&$K$(MeV)&$k_{eq}$(MeV)&$B$(MeV)\\ \hline
2.3&1960&313&50.4\\ \hline
2.4&1275&308&37.0\\ \hline
2.5&687&297&27.1\\ \hline
2.6&309&279&20.4\\ \hline
2.7&196&257&16.4\\ \hline
2.8&184&241&14.0\\ \hline
2.9&180&231&12.4\\ \hline
3.0&175&223&11.2\\ \hline
3.1&169&217&10.3\\ \hline
\end{tabular}
\end{center}
\end{table}

\subsubsection{Many-body forces}
\begin{figure}[bthp]
\setlength{\epsfysize}{8in}
\centerline{\epsffile{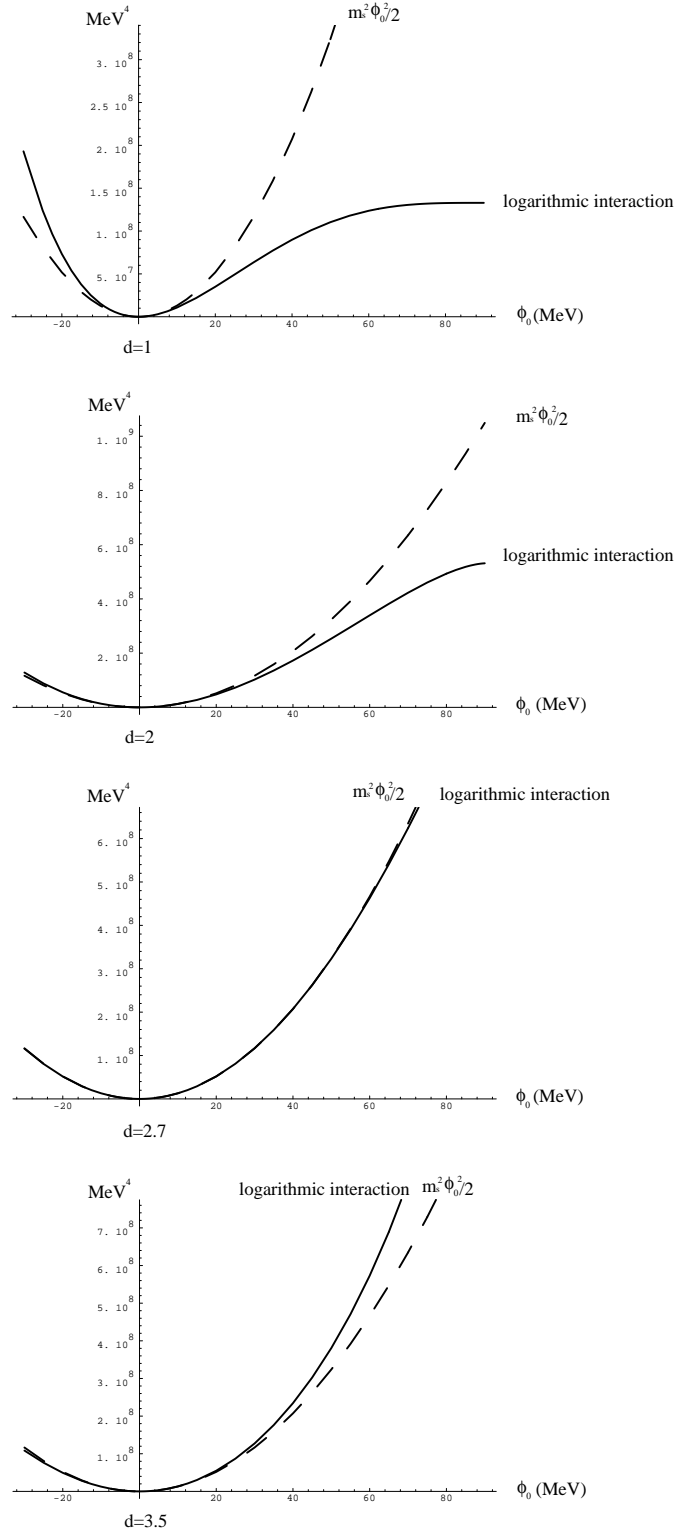}}
\caption[]{\small
Comparison between the $\phi^2$ interaction and the logarithmic 
self-interaction of the scalar field { 
with the FTS1 parameters.} The dashed lines represent $
\frac{m_s^2}{2}\phi^2$ and the solid lines 
$\frac{m_s^2}{4}S_0^2d^2[(1-\frac{\phi}{S_0})^{4/d}[\frac{1}{d}
ln(1-\frac{\phi}{S_0})-\frac{1}{4}]+\frac{1}{4}]$ for (from top to
bottom) $d=1,2,2.7,3.5$  respectively.}\label{pot}
\end{figure}

\indent\indent
In mean field, the logarithmic potential in Eqs. (\ref{leff}) and 
(\ref{energydensity})
contains n-body-force (for $n\geq 2$) contributions to the energy density.
For the FTS1 parameters, these n-body terms are strongly suppressed
for $d\gsim 2.6$. This is shown in Figure \ref{pot} where it is seen that
the entire potential term is accurately reproduced by the quadratic term
$\frac 12 m_s^2\phi^2$ for $d_{an}\sim 5/3$.  
Furthermore a close examination
of the results reveals that each of the n-body terms are separately
suppressed. This phenomenon is in some sense consistent with chiral symmetry
\cite{weinberg} and is observed in the spectroscopy of light nuclei 
\cite{friar}.

\begin{figure}
\setlength{\epsfysize}{3in}
\setlength{\epsfxsize}{4in}
\centerline{\epsffile{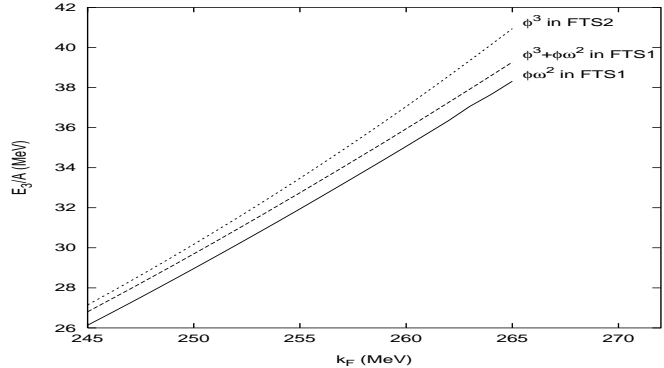}}
\caption[]{\small The 3-body contributions to the energy per nucleon
vs. Fermi momentum in FTS models.
The short-dashed line represents the contribution of the $\phi^3$ term
in FTS2 with the Q1 parameters. The long-dashed and the solid lines represent 
the contributions of the cubic terms ($\phi
\omega^2$ and $\phi^3$) in FTS1 
with the T1 parameters for $d=2.7$.}\label{three}
\end{figure}

Since there are additional polynomial terms in vector fields
(i.e., terms like $\phi\omega^2$), 
the near complete suppression of the scalar polynomials
does not mean the same for the total many-body forces.
In fact we should not expect it. To explain why this is so, we plot in 
Figure \ref{three} the
three-body contributions of the $\phi^3$ and $\phi\omega^2$ forms.
We also compare the FTS1 results with the FTS2 \cite{tang2} results that are based on the naturalness condition. In FTS1, the $\phi^3$ term which
turns to repulsion from attraction for $d>8/3$ contributes little, so the main repulsion arises from the $\phi\omega^2$-type term. This, together with
an attraction from a $\omega^4$ term, 
is needed for saturation of the nuclear matter at the right density\footnote{
This raises the question as to how
one can understand the result obtained by Brown, Buballa and Rho \cite{BBR}
where it is argued that the chiral phase transition in dense medium
is of mean field with
the bag constant given entirely
by the quadratic term $\sim \frac 12 m_s^2\phi^2$.
The answer to this question is as follows. First we expect that
the anomalous dimension will stay locked at $d_{an}=d-1\sim 5/3$ near the
phase transition (this is because the anomalous dimension associated with
the trace anomaly -- a consequence of ultraviolet regularization --
is not expected to depend upon density), 
so the $\phi^n$ terms for $n>2$ will continue
to be suppressed as density approaches the critical value. Secondly near the chiral phase transition,  the gauge coupling of the vector mesons,
as argued in \cite{BRPR96}, will
go to zero in accordance with the Georgi vector limit \cite{vector}, so
the many-body forces associated with the vector mesons will also be 
suppressed.}.

\subsection{Anomalous dimension and the scalar-meson mass}\label{dim}
\indent\indent
We would like to understand how the large anomalous 
dimension needed here could
arise in the theory and its role in the scalar sector. 
Since the trace anomaly arises from the 
necessity to regularize the theory in the ultraviolet, 
it cannot depend on density as long as the
Fermi momentum involved is less than the chiral scale $\Lambda_\chi$. Thus
the anomalous dimension cannot be due to an effect of density on the trace
anomaly. This means instead that the large anomalous dimension reflects
a strong-coupling regime with 
the fluctuation around the matter-free vacuum being strong.
 
As suggested in \cite{FR96} and elaborated more in the next section, 
one appealing way of 
understanding the FTS1 mean field theory is to consider {\it all channels} to be at Fermi liquid fixed points except that because of trace anomaly, the scalar field develops an anomalous dimension, thereby affecting the 
four-Fermi interaction in the scalar channel resulting when the scalar
field is integrated out. 
If the anomalous dimension were 
sufficiently negative so that marginal terms became marginally relevant, then
the system would become unstable as in the case of the NJL model or 
superconductivity, 
with the resulting spontaneous
symmetry breaking. However if the anomalous dimension is positive, then 
the resulting effect will instead be a screening. The positive anomalous
dimension we need here belongs to the latter case. We can see this as follows.
Consider the potential given with the low-lying scalar
$S$ (with the gluonium component integrated out):
\be
V(S,\cdots)=\frac 14 m_S^2 d^2 S_0^2 \left(\frac{S}{S_0}\right)^{\frac 2d}
\left(\frac 1d \ln\frac{S}{S_0} -\frac 14\right) +\cdots\label{V}
\ee
where $m_S$ is the light-quarkonium mass in free space ($\sim 700$ MeV) and
the ellipses stand for other contributions such as pions, quark masses etc. that
we are not interested in. The scalar excitation on a given background
$S^\star$ is given by the double derivative of $V$ with respect to $S$
at $S=S^\star$
\be
{m_S^\star}^2=m_S^2 \left(\frac{S^\star}{S_0}\right)^{
\frac 4d-2}\left[1+\left(\frac 4d -1\right)\ln\frac{S^\star}{S_0}\right].
\ee
We may identify the ratio $S^\star/S_0$ with the BR scaling factor $\Phi$
\cite{FR96}:
\be
\frac{S^\star}{S_0}=\Phi=\frac{f_\pi^\star}{f_\pi}=\frac{m_V^\star}{m_V}
\ee
with the subscript V standing for light-quark vector mesons $\rho$
and $\omega$.  Then we have
\be
\frac{m_S^\star}{m_S}=\Phi (\rho) \kappa_d (\rho)
\ee
with
\be
\kappa_d (\rho)= \Phi^{\frac 2d -2} \left[1+(\frac 4d -1)
\ln\Phi\right]^{\frac 12}.
\ee
One can see that for $d=1$ which would correspond to the canonical dimension
of a scalar field the scalar mass falls much faster, for a $\Phi (\rho)$
that decreases as a function of density, than what would be given
by BR scaling. Increasing the $d$ (and hence the anomalous dimension) makes the
scalar mass fall less rapidly. With $d\approx 2$, $\kappa_d \approx 1$
and we recover the BR scaling. Since the dropping scalar mass is associated
with an increasing attraction, we see that the anomalous dimension plays
the role of bringing in an effective repulsion. One may therefore interpret this
as a screening effect of the scalar attraction. In particular, that
$d-2\approx .7 >0$ means that in FTS1, an additional screening of the
BR scaled scalar exchange (or an effective repulsion) is present.

\section{Chiral Liquid and Fermi-Liquid Fixed Point}
\indent\indent
In a more recent paper, Furnstahl, Serot and Tang \cite{tang2} 
reformulated their
theory in terms of a chiral Lagrangian constructed by
applying the ``naturalness"
condition for all relevant fields. In doing this, Georgi's ``naive
dimensional analysis" \cite{NDA} 
was used instead of the trace anomaly and the large anomalous
dimension. It was argued therein that a Lagrangian so
constructed contains in principle higher-order terms in chiral counting 
including those loop corrections that can be expressed as
counter terms involving matter fields (e.g., baryons). 
This is essentially equivalent to
Lynn's effective action \cite{lynn} that purports to include all orders of 
quantum loops in chiral expansion supplemented with counter terms 
consistent with the order to which loops are calculated. This means
that the mean-field solution with the FTS1 (or equivalently \cite{tang2})
should correspond to the ``chiral liquid"  as the ground-state
matter that arises as a non-topological soliton proposed by Lynn.
Fluctuations around this mean fields should then give an accurate
description of the observables that we are dealing with.

We shall here extend this argument further and make a contact with
Landau's Fermi-liquid theory of nuclear matter by using the argument
of Matsui \cite{matsui} who described the link between Walecka model
in mean field and Landau-Migdal Fermi liquid theory. This will
allow us to understand BR scaling in terms of chiral Lagrangians
and Fermi-liquid fixed point theory thereby giving a unified picture
of ordinary nuclear matter and extreme state of matter probed in
heavy-ion collisions, e.g., CERES. As far as we know this is the
first such attempt to connect the physics of the two vastly different
regimes. The seed for such a scheme and the basic idea were mentioned
in the work of Friman and Rho \cite{FR96}.

The basic assumption we start with is that the chiral liquid arises
from a quantum effective action resulting from integrating out the
degrees of freedom lying above the chiral scale $\Lambda_\chi
\sim 4\pi f_\pi \sim 1$ GeV. This corresponds to the first stage
of ``decimation''\cite{froehlich} in our scheme.
The mean field solution of this action is then supposed 
to yield the ground state of nuclear matter with the Fermi surface
characterized by the Fermi momentum $k_F$. In FTS1, the effective 
Lagrangian was given in terms of the baryon, pion, quarkonium scalar and
vector fields with the gluonium scalars integrated out.
Instead of
treating the scalar and vector fields {\it explicitly} as in FTS1,
we will consider here integrating them out further from 
the effective Lagrangian.
This would lead to four-Fermi, six-Fermi etc. interactions in the 
Lagrangian with various powers of derivatives acting on the Fermi field.
The resulting effective Lagrangian will then consist of the baryons and
pions coupled bilinearly in the baryon field and four-Fermi and
higher-Fermi interactions with various powers of derivatives, all 
consistent with chiral symmetry. A minimum version of such Lagrangian
in mean field can be shown to lead to the original (naive) Walecka model
\cite{tsp}. In principle a sophisticated version of this procedure\
should give a theory equivalent to the full
FTS1 theory or a generalization thereof.

Leaving out the pion for the moment\footnote{The pion will be introduced 
in the next section in terms of a non-local four-Fermi interaction 
that enters in the ground state property and gives the nucleon Landau mass
formula in terms of BR scaling and pionic Fock term. See later.} 
and formulated 
non-relativistically\footnote{One could do this relativistically as
shown by Baym and Chin\cite{baymchin} which will be necessary for
heavy-ion collisions but we will present the arguments in non-relativistic
form.}, the next step is to decimate successively
the degrees of freedom present in the excitations
with the scale $E<\Lambda_\chi$
as follows\footnote{Here we are relying on the procedure of ``decimation"
formulated rigorously by
Fr\"ohlich et al\cite{froehlich} in condensed matter physics.}.
To do this, we consider excitations near the  
Fermi surface which we shall take to be spherical for convenience
characterized by
$k_F$. First integrate out the excitations with momentum 
$p\geq \pm \Lambda$ (where $p=|\vec{p}|$ and $\Lambda <\Lambda_\chi$) 
measured relative to 
$k_F$ (corresponding to the
particle-hole excitations with momentum greater than $2\Lambda$). We 
are thus restricting ourselves to the physics of
excitations whose momenta lie
below $2\Lambda$.
This defines the starting point of an {\it in-medium}
renormalization group procedure.
The appropriate action to consider can be written in a simplified 
and schematic form as
\be
S=\int_\Lambda \bar{\psi}[i\omega-v_F^\star k]\psi +\delta\mu^\star
\int_\Lambda \bar{\psi}\psi +\int_\Lambda u\bar{\psi}\bar{\psi}\psi\psi
\label{Flag}
\ee
where
\be
\int_\Lambda:=\int \frac{d\Omega}{(2\pi)^2}\int_{-\Lambda}^\Lambda
\frac{dk}{(2\pi)}\int_{-\infty}^\infty \frac{d\omega}{(2\pi)}.
\ee
Here $v_F^\star=k_F/m^\star$ where $m^\star$ is the effective mass of the
nucleon which will be equal to the Landau mass $m_L^\star$ as will be
elaborated on later. The term with $\delta \mu^\star$ is a counter term added
to assure that the Fermi momentum is fixed (that is, the density is fixed).
What his term does is to cancel loop contributions involving the four-Fermi
interaction to the nucleon self-energy (i.e., the ``tadpole'') so that
the $v_F^\star$ is at the fixed point. This means that the counter term
essentially assures that the effective mass $m^\star$ be at the fixed point.
Without this procedure, the term quadratic in the fermion field would be
``relevant'' and hence would be unnatural \cite{shankar}.

In nuclear matter, the spin and isospin degrees of freedom
need to be taken into account into the four-Fermi interaction. We have
written all these symbolically in the action (\ref{Flag}). The function
$u$ in the four-Fermi interaction term can therefore contain spin and isospin 
factors as well as space dependence that takes into account non-locality 
and derivatives. For simplicity we will consider it to be a constant
depending in general on spin and isospin factors. 
Non-constant terms will be ``irrelevant.'' We shall ignore in the next sections
the spin dependence which will be considered elsewhere, thus confining
ourselves to the Landau parameters $F$ and $F^\prime$ corresponding to
the particle-hole vibrational channel. 
In our discussions, the BCS channel that corresponds to a particle-particle 
channel does not figure and hence will not be considered 
explicitly.

The upshot of the analyses in \cite{shankar} and \cite{froehlich} 
which we apply to our system is that
in addition to the Fermi surface fixed point with the $m^\star$, 
the four-Fermi interactions in the phonon channel $F$ are also
at the fixed points. 
In general four-Fermi interactions are irrelevant except for special 
kinematics for which the interaction becomes marginal leading to fixed
points. Six-Fermi and higher-Fermi interactions are always irrelevant
and can contribute at most to screening of the fixed-point constants.
{\it Since the parameters of the fixed-point theory are taken from experiments,
we need not worry about this renormalization}. The resulting theory
is the Fermi-liquid fixed point theory.
Shankar arrives at this theory by showing, 
in the absence of BCS interactions, that in the large $N$ limit
where $1/N=\Lambda/k_F$, only
one loop contributions survive. 
Fr\"ohlich et al obtain the same result in the $1/N$ expansion where their $N$ 
is taken to be $N\sim \lambda$ with $1/\lambda$ being the width of the
effective wave vector space around the Fermi sea which can be considered as
the ratio of the microscopic scale to the mesoscopic scale. 
More specifically if one rescales the four-Fermi interaction such that
one defines the dimensionless constant $g$,
$u_0\sim g/k_F^2$ where $u_0$ is the 
leading term (i.e., constant term) in the Taylor series of the quantity
$u$ in (\ref{Flag}), then the fermion wave function renormalization
$Z$, the Fermi velocity $v_F$  and the constant $g$
are found not to flow up to order ${\O}(g^2/N)$. Thus in the 
large $N$ limit, the
system flows to Landau fixed point theory to all orders of loop corrections.
This result is correct provided there are no long-range interactions and
if the BCS channel is turned off. One can show this also in terms of 
bosonization which turns out to be possible because of dimensional
reduction of the Fermi liquid system to an effective one-dimensional 
Dirac system as shown in \cite{froehlich2}. 

In sum, we arrive at the picture where the chiral liquid solution of
the quantum effective chiral action gives the Fermi liquid fixed point
theory. The parameters of the four-Fermi interactions in the phonon
channel are then identified with the fixed-point Landau parameters.
This identification would allow the mapping of the BR scaled parameters
to the quantities governed by the Landau parameters $F$ and $F^\prime$
discussed in the following subsection.

\section{BR Scaling and More Effective Chiral Lagrangians}
\subsection{The power of BR scaling}
\indent\indent
If the large anomalous dimension of the scalar field in FTS1 is a symptom
of a strong-coupling regime, it suggests that one should redefine the
vacuum in such a way that the fluctuation around the new vacuum becomes
weak-coupling. This is the basis of the BR scaling introduced in \cite{BR91}.
The basic idea\footnote{For completeness, we briefly summarize the key 
argument of \cite{BR91}. Consider an extended chunk of nuclear matter.
If the system is sufficiently dilute, one can start with a 
chiral Lagrangian constructed with
parameters fixed in the matter-free space
characterized by a corresponding scale, say, $\Lambda_0$. 
Now suppose that the matter is ``squeezed" to a  
density $\rho$ with its scale characterized by,  say, $\Lambda_\rho$.
Our basic assumption is that to describe this dense system, we may impose 
the same symmetry (such as chiral symmetry, conformal anomaly etc.)
constraints as in the matter-free space while replacing in the effective 
chiral Lagrangian the
free-space parameters -- masses and coupling constants -- by those defined
at that density. BR scaling is one specific way of defining these
modified parameters.} is to fluctuate around the ``vacuum" defined at
$\rho\approx
\rho_0$ characterized by the quark condensate $\la \bar{q}q\ra_{\rho}
\equiv \la\bar{q}q\ra^\star$.
In \cite{BR91,BRPR96}, 
this theory was developed with a chiral Lagrangian implemented
with the trace anomaly of QCD. The Lagrangian used was the one valid in the 
large
$N_c$ limit of QCD and hence given entirely in terms of boson fields from 
which baryons arose as solitons (skyrmions): Baryon properties are therefore
dictated by the structure of the bosonic Lagrangian, thereby leading
to a sort of {\it universal scaling} between mesons and baryons. One can see
using a dilated chiral quark model that the BR scaling is a generic
feature also at high temperature in the large $N_c$ limit\cite{hkl}.

In this description, one is approximating the complicated strong interaction
process at a nuclear matter density in terms of ``quasiparticle" 
excitations for 
both baryons and bosons in medium. This means that the properties of 
fermions and bosons in medium at $\rho\approx \rho_0$ are given in terms of 
tree diagrams with the properties defined in terms of the masses and coupling 
constants universally determined by the quark condensates at that density.

The question then is: How can one ``marry" the FTS1 Lagrangian with
the BR scaled Lagrangian? The next question is how to identify BR scaled
parameters with the Landau parameters. In the rest of this section,
we will provide some answers to these two questions.

\subsection{A hybrid model} 
\indent\indent
As a first attempt to answer this question, we consider the hybrid
model in which the ground state is given by the mean field of the FTS1
Lagrangian ${\L}_{FTS1}$ 
and the fluctuation around the ground state is described
by the tree diagrams of the BR scaled Lagrangian $\Delta {\L}$,
\be
{\L}^{eff}={\L}_{FTS1} + \Delta {\L}.
\ee
Note that the fluctuation in the strangeness direction (\ref{kaonL})
discussed below corresponds to one of the terms figuring in $\Delta {\L}$.
This model with the canonical parameters (T1) for the ground state
and a BR scaled fluctuation Lagrangian
in the non-strange flavor sector was recently used by Li,
Brown, Lee and Ko \cite{LBLK96} for describing {\it simultaneously}
nucleon flow and dilepton production in heavy-ion collisions.
The nucleon flow is sensitive to the parameters of the baryon sector, in particular the repulsive
nucleon vector potential at high density whereas
the dilepton production probes the parameters of the meson sector.
With a suitable momentum dependence implemented to the FTS1 mean field
equation of state, the nucleon flow comes out in good agreement with
experiments. Furthermore the scaling of the nucleon mass in the 
FTS1 theory in dense medium, say, at $\rho\sim 3\rho_0$, is found to
be essentially the same as that given by the NJL model. Therefore 
we can conclude that the nucleon in FTS1 scales in the same way
as BR scaling. 

The dilepton production involves both baryon and meson
properties, the former in the scaling of the nucleon mass and the latter
in the scaling of the vector meson ($\rho$) mass. The equation of state
correctly describing the nucleon flow and the BR scaled vector meson mass
are found to fit the dilepton data equally well, comparable
to the fit obtained in
\cite{LKB} using  Walecka mean field. What is important in this process is
the scalar mean field which governs the BR scaling and hence the
production rate comes out essentially the same for FTS1 and Walecka
mean fields. The delicate interplay between the attraction and the
repulsion that figures importantly  in the compression modulus \cite{SBMR2}
does not play
 an important role in the dilepton process.

Let us see how the particles behave in the background of the FTS1 ground
state given by ${\L}_{FTS1}$. The nucleon of course scales \`a la BR as
mentioned above. We can say nothing on the pion and the
$\rho$ meson with the FTS1 theory. However there is nothing which would
preclude the $\rho$ scaling \`a la BR and the pion non-scaling
within the scheme. What is encoded in the FTS1 theory is the behavior
of the $\omega$ and the scalar $S$ which figure importantly in Walecka
mean fields. Let us therefore focus on these two fields in medium near 
normal nuclear matter density.

We have already shown in subsection \ref{dim} that the mass of the 
scalar field $S$ drops less rapidly than BR scaling for $d>2$. One can 
think of this as a screening of the four-Fermi interaction in the scalar 
channel that arises when the scalar meson
with the BR scaled mass is integrated out.  This feature and the property
of the $\omega$ field can be seen by the toy model of the FTS1 Lagrangian
(that includes terms corresponding up to three-body forces)
\be
\L_{toyFTS1}=\L_{BR}
 +\frac{m_\omega^2}{2}(2+\eta )\frac{\phi}{S_0}\omega^2 -\frac{m_s^2\phi^3}
{3S_0}\label{toy}
\ee
where
\be
\L_{BR} &=& \bar{N}(i\gamma_{\mu}(\d^\mu+ig_v\omega^\mu )-M+g_s\phi )N
\nonumber\\
& &+\frac{m_\omega^2}{2}\omega^2 (1-\frac{2\phi}{S_0})
-\frac{m_s^2}{2}\phi^2 (1-\frac{2\phi}{3S_0}).\label{toyBR}
\ee
We have written ${\L_{BR}}$ such that the BR scaling is incorporated {\it at
mean field level}
as\footnote{Here we are ignoring the deviation of the scaling of the 
effective nucleon mass (denoted later as $m_L^\star$) \cite{FR96} from the
universal scaling $\Phi (\rho)$. This will be incorporated in the next
subsection.}
\be
\Phi (\rho) = \frac{M^\star}{M}=\frac{m_s^\star}{m_s}
=\frac{m_\omega^\star}{m_\omega}\approx 1-\frac{\phi}{S_0}
\ee
with
\be
S_0=\la 0|S|0\ra = M/g_s.
\ee
We can see from (\ref{toy}) that the FTS1 theory brings in an additional
repulsive three-body force  coming from a cubic scalar field term
while if one takes $\eta=-2$, the $\omega$ field will have a BR scaling mass
in nuclear matter. Fit to experiments favors $\eta\approx -1/2$ instead 
of $-2$, thus
indicating that the FTS1 theory requires a many-body suppression of the
repulsion due to the $\omega$ exchange two-body force. (In the simple model
with BR scaling that we will construct below, we shall use this feature by
introducing a ``running'' vector coupling $g_v^\star$ that drops as a function 
of density.)
The effective $\omega$ mass may be written as
\be
m_\omega^{\star 2}\approx [1+\eta\frac{\phi_0}{S_0}]m_\omega^2.
\ee
For $\eta <0$, we have a falling $\omega$ mass corresponding to
BR scaling (modulo, of course, the numerical value of $\eta$). In 
FTS1, there is a quartic term $\sim \omega^4$ which is attractive and
hence {\it increases} the $\omega$ mass. In fact,  
because of the attractive
quartic $\omega$ term, we have
\be
\frac{m_\omega^\star}{m_\omega}\approx1.12\label{omegaeff}
\ee
at the saturation density with T1 parameter set. 
This would seem to suggest that due to higher
polynomial (many-body) effects,
the $\omega$ mass does not follow BR scaling in medium.
Furthermore the $\omega$ effective mass 
increases slowly  
around this equilibrium value:
\be
\frac{\d m_\omega^\star}{\d k_F}
\sim \frac{0.0004}{\mbox{MeV}^2}\ \alpha
\ee
with $\alpha\equiv\frac{\gamma}{2\pi^2}k_F^2$, 
if one uses
\be
\phi_0
&\approx&\frac{g_s}{m_s^2}\frac{\gamma}{6\pi^2}k_F^3\\
\omega_0
&\approx&\frac{g_v}{m_\omega^2}\frac{\gamma}{6\pi^2}k_F^3
\ee
with the degeneracy factor $\gamma$ and the T1 parameters.
\subsection{Model with BR scaling}\label{model}
\indent\indent
The above hybrid model suggests how to construct an effective Lagrangian
model that implements BR scaling and contains the same physics as 
FTS1 theory. The crucial point is that such a 
Lagrangian is to give {in mean field} the chiral
liquid soliton solution. This can be done by making
the following replacements in (\ref{toyBR}):
\be
M-g_s\phi_0&\rightarrow& M^\star,\nonumber\\
m_\omega^2 (1-\frac{2\phi_0}{S_0})&\rightarrow& {m_\omega^\star}^2,\nonumber\\
m_s^2 (1-\frac{2\phi_0}{S_0})&\rightarrow& {m_s^\star}^2
\ee
and write
\be
\L_{BR} &=& \bar{N}(i\gamma_{\mu}(\d^\mu+ig_v\omega^\mu )-M^\star+h\phi )N
\nonumber\\
& &-\frac 14 F_{\mu\nu}^2 +\frac 12 (\partial_\mu \phi)^2 
+\frac{{m^\star_\omega}^2}{2}\omega^2 
-\frac{{m^\star_s}^2}{2}\phi^2\label{toyBRP}
\ee
with
\be
\frac{M^\star}{M}=\frac{m_\omega^\star}{m_\omega}=\frac{m_s^\star}{m_s}
=\Phi (\rho).
\ee
The additional term $\bar{N}h\phi N$ is put in to account for the difference
between the Landau mass $m_L^\star$ to be given later and the BR scaling
mass $M^\star$. In the chiral Lagrangian approach with BR scaling, the
difference comes through the Fock term involving non-local pion 
exchange\cite{FR96}. This will be discussed further in the next subsection. 
For simplicity we will take the scaling in the form
\be
\Phi (\rho)=\frac{1}{1+y\rho/\rho_0}
\ee
with $y=0.28$ so as to give $\Phi (\rho_0)=0.78$ (corresponding to
$k_F=260$ MeV) found in QCD sum-rule calculations 
\cite{FR96} as will be discussed shortly, 
as well as from the {\it in-medium} Gell-Mann-Oakes-Renner relation 
\cite{BRPR96}. 

Note that the Lagrangian (\ref{toyBRP}) treated at mean field level would 
give a Walecka-type model
with the meson masses replaced by the BR scaling mass. 

Now in order to describe
nuclear matter in the spirit of the FTS1 theory, we  introduce
terms cubic and higher in $\omega$ and $\phi$ fields {\it
to be treated as perturbations around the BR background} as
\be
\L_{n-body}=a \phi\omega^2 + b \phi^3 +c \omega^4 +d \phi^4 +e \phi^2\omega^2
+\cdots\label{nbody}
\ee
where $a$ -- $e$  are ``natural" (possibly density-dependent) 
constants to be 
determined. By inserting for 
the $\phi$ and $\omega$ fields the solutions 
of the static mean field equations given by ${{\cal L}_{BR}}$, 
\be
{m_s^\star}^2\phi &=&h\sum_i\bar{N}_iN_i\label{meanphi}
\\
{m^\star_\omega}^2\omega &=&g_v\sum_i N_i^\dagger N_i
\ee
 we see that at mean-field level,
${\L_{n-body}}$ generates three- and higher-body forces with 
the exchanged masses density-dependent \`a la BR. {\it Note that at this point,
the scaling factor $\Phi$ and the mean field value (\ref{meanphi})
are not necessarily locked to each other by self-consistency.}

As the first trial, we will consider the drastically simplified model
by dropping the n-body term (\ref{nbody}) and minimally
modifying the BR Lagrangian (\ref{toyBRP}). We shall do this
by letting as mentioned above 
the vector coupling ``run'' as a function of density.
For this,  we use the observation made in 
\cite{LBLK96} that the nucleon flow probing higher density requires that
$g_v^\star/m_v^\star$ be independent of density at low densities and
decrease slightly at high densities. We shall therefore take, to simulate
this particular many-body correlation effect, the vector 
coupling to scale as
\be
\frac{g_v^\star}{g_v}=\frac{1}{1+z\rho/\rho_0}
\ee 
with $z$ equal to or slightly greater than $y$.\footnote{This scaling seems
at odds with the prediction made with the Skyrme model \cite{skyrmescaling}
where using the Skyrme model with the quartic Skyrme term inversely
proportional to the coupling $e$, it was found that
\be
\frac{e}{e^\star}\sim \sqrt{\frac{g_A^\star}{g_A}}.\nonumber
\ee
It is tempting to identify (via $SU(6)$ symmetry) $e$ with $g_v$ 
that we are discussing here
since the Skyrme quartic term
can formally be obtained from a hidden gauge-symmetric Lagrangian by 
integrating out the $\rho$ meson field. If this were correct, one would
predict that the vector coupling increases -- and not decreases --
as density increases since
we know that $g_A^\star$ is quenched in dense matter. This identification
could be too naive and incomplete in two respects, however.  
First of all, this skyrmion formula is a large-$N_c$ relation
and secondly the Skyrme quartic term embodies {\it all} short-distance physics 
in one dimension-four term
in a derivative expansion. Thus the constant $1/e$ must represent a lot
more than just the vector-meson ($\rho$) degree of freedom. 
Furthermore we are concerned with the $\omega$ degree of freedom which
in a naive derivative expansion would give a six-derivative term.
The BR scaled model we are constructing should involve not only 
short-distance physics presumably represented
by the $1/e$  term (consistent with the understanding
that the quenching of $g_A$ is a short-distance phenomena) but also 
longer-range correlations. Therefore the qualitative difference should 
surprise no one.}

The truncated
Lagrangian that we shall consider then is
\be
\L_{BR} &=& \bar{N}(i\gamma_{\mu}(\d^\mu+ig_v^\star\omega^\mu )-M^\star+h\phi )N
\nonumber\\
& &-\frac 14 F_{\mu\nu}^2 +\frac 12 (\partial_\mu \phi)^2 
+\frac{{m^\star_\omega}^2}{2}\omega^2 
-\frac{{m^\star_s}^2}{2}\phi^2.\label{toyBRPP}
\ee

In Table \ref{yeqz}, three sets of parameters are listed. We take
the measured free-space masses for the $\omega$ and the nucleon and
for the scalar $\phi$ for which the free-space mass cannot be precisely
given, we take $m_s=700$ MeV (consistent with what is argued in \cite{BRPR96})
so that at nuclear matter density, it comes close to what enters in the FTS1.

\begin{table}
\caption{ Parameters for the Lagrangian (\ref{toyBRPP})  with $y=0.28, 
m_s=700{\mbox{MeV}}, m_\omega =783 
{\mbox{MeV}}, M=939{\mbox{MeV}}$}\label{yeqz}
\vskip .3cm
\begin{center}
\begin{tabular}{|c|c|c|c|}\hline
SET&$h$& $g_v$&$z$\\ \hline
S1&6.62&15.8&0.28 \\ \hline
S2&5.62&15.3&0.30 \\ \hline
S3&5.30&15.2&0.31 \\ \hline
\end{tabular}
\end{center}
\end{table}
The resulting fits to the properties of nuclear matter are given in
Table \ref{ryez} for the parameters given in Table \ref{yeqz}.
\begin{table}
\caption{Nuclear matter properties predicted with the parameters of Table 
\ref{yeqz}. The effective nucleon mass (later identified with the
Landau mass) is $m_L^\star=M^\star-h\phi_0$.}\label{ryez}
\vskip .3cm
\begin{center}
\begin{tabular}{|c|c|c|c|c|c|}\hline
SET&$E/A-M$(MeV)&$k_{eq}$(MeV)&K(MeV)& $m_L^\star /M$&$\Phi (k_{eq})$\\ \hline
S1&-16.0&257.3&296&0.619&0.79 \\ \hline
S2&-16.2&256.9&263&0.666&0.79 \\ \hline
S3&-16.1&258.2&259&0.675&0.78 \\ \hline
\end{tabular}
\end{center}
\end{table}

These results are encouraging. Considering
the simplicity of the model, the model -- in particular with the S2 and 
S3 set --
is remarkably
close in nuclear matter to the full FTS1. 
The compression modulus comes down toward 
the low value that is currently favored.  In fact, the somewhat higher 
value obtained here can be easily brought down to about 200 MeV
without modifying other quantities if one admits a small admixture of 
the {\it residual} many-body terms (\ref{nbody}), as we shall shortly show.
The effective nucleon Landau mass
$m_L^\star/M\approx 0.67$ is in good 
agreement with what was obtained in QCD sum-rule calculations (see
\cite{FR96})
and also below (i.e., 0.69)
by mapping BR scaling to Landau-Migdal Fermi liquid theory. We shall see below
that this has strong support from low-energy nuclear properties.
What is also noteworthy is that the ratio
${\cal R}\equiv (g_v^\star/m_v^\star)^2$ forced upon us -- though not predicted
-- is independent of the density
(set S1) or slightly decreasing with density (sets S2 and S3),
as required in the nucleon flow data as found by
Li, Brown, Lee and Ko\cite{LBLK96}\footnote{In FTS1 theory, it is the higher
polynomial terms in $\omega$ and $\phi$ defining the mean fields that
are responsible for the reduction in ${{\cal R}}$ needed in \cite{LBLK96}.
In Dirac-Brueckner-Hartree-Fock theory, it is found \cite{brockmann} that while
${{\cal R}}$ takes the free-space value ${{\cal R}}_0$ for 
$\rho\approx \rho_0$, it decreases to
${{\cal R}}\approx 0.64{{\cal R}}_0$ 
at $\rho \approx 3\rho_0$ due to rescattering
terms which in our language would correspond to the many-body correlations. 
}.

\begin{figure}[tbh]
\setlength{\epsfysize}{4.0in}
\centerline{\epsffile{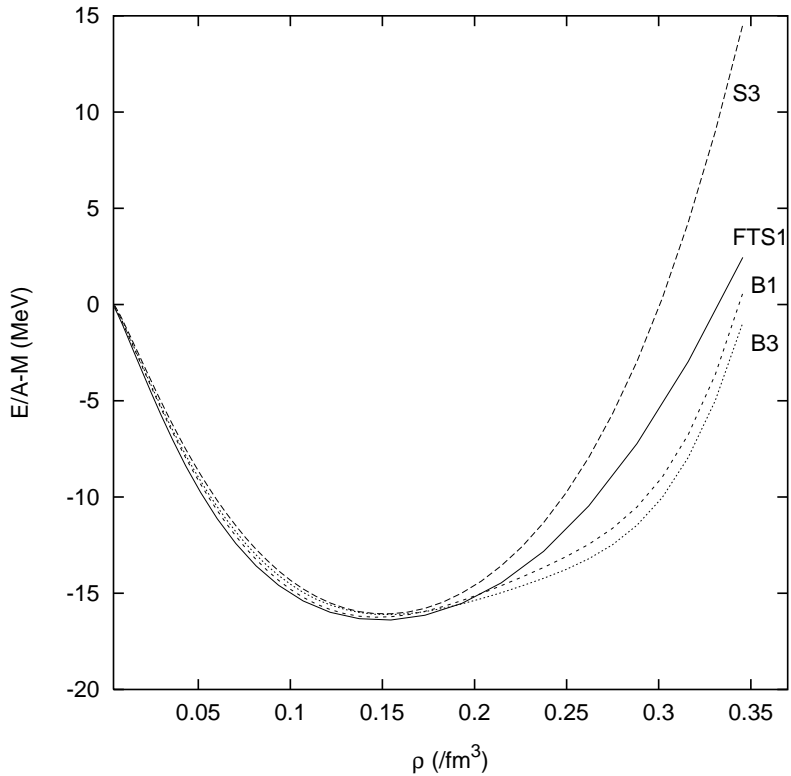}}
\caption[good]{$E/A-M$ vs. $\rho$ 
for FTS1 theory (``T1'' parameter), the ``S3'',
``B1'' and ``B3'' models 
defined in Table \ref{fit}.}\label{good}
\end{figure}

The assumption that the many-body correlation terms in (\ref{nbody}) can be
entirely subsumed in the dropping vector coupling may seem too drastic. Let us
see what small residual
three-body and four-body terms in (\ref{nbody}) as many-body correlations
(over and above what is 
included in the running vector coupling constant) can do to 
nuclear matter properties. For convenience we rewrite (\ref{nbody}) by
inserting dimensional factors as 
\be
\L_{n-body}&=&\frac{\eta_0}{2}m^2_\omega\frac{\phi}{f_\pi}\omega^2 
-\frac{\kappa_3}{3!} m_s^2\frac{\phi^3}{f_\pi} 
\label{nbody2}\\
& &+\frac{\zeta_0}{4!}g_v^2\omega^4
-\frac{\kappa_4}{4!}m_s^2\frac{\phi^4}{f_\pi^2}
+\frac{\eta_1}{2}m^2_\omega\frac{\phi^2}{f_\pi^2}\omega^2
\nonumber
\ee 
and demand that the coefficients $\eta$, $\zeta$ and $\kappa$ so defined 
be natural. The results of this analysis are given in Table \ref{fit} and
Figure \ref{good} 
for various values of
the residual many-body terms and compared with those of the truncated model
(\ref{toyBRPP}) with S3 parameter set.
The coefficients are chosen somewhat arbitrarily to bring our 
points home, with no attempt made for
a systematic fit. (It would be easy to fine-tune the parameters to
make the model as close as one wishes to FTS1 theory.)
It should be noted that
while the equilibrium density or Fermi momentum $k_{eq}$, 
the effective nucleon mass $m_L^\star$ and
the binding energy $B$ stay more or less
unchanged, within the range of the parameters chosen, from what is given 
by the BR-scaled model (\ref{toyBRPP}) with the S3 parameters,
the compression modulus $K$ can be substantially decreased by the residual
many-body terms. Figure \ref{good} shows that as expected, lowering 
of the compression modulus is accompanied by softening of the equation
of state at $\rho > \rho_0$. While the equilibrium property other than
the compression modulus is insensitive to the many-body correlation terms,
the EOS at larger density can be quite sensitive to them.
This is because for the generic parameters chosen,  
the $m_L^\star$ can vanish 
at a given density above $\rho_0$ at which the 
approximation
is expected to break down and hence the resulting result cannot be trusted.
The B2 and B4 models do show this instability {at $\rho\gsim 1.5\rho_0$}. 
(See Figure \ref{comparison} in the Appendix.) 

\begin{table}
\caption{Effect of many-body correlations on nuclear matter
properties using the Lagrangian  
(\ref{toyBRPP}) $+$ (\ref{nbody2}). We have fixed the free-space masses
 $m_s=700{\mbox{MeV}}, m_\omega =783 {\mbox{MeV}}, M=939{\mbox{MeV}}$ and
set $\eta_1=0$ for simplicity. The equilibrium density
$k_{eq}$, the compression modulus $K$, and the binding energy 
$B=M-E/A$ are all given in units of MeV.}
\label{fit}
\vskip .3cm
{\small
\begin{tabular}{|c|c|c|c|c|c|c|c|c||c|c|c|c|}\hline
SET&$h$&$g_v$&$y$&$z$&$\eta_0$&$\zeta_0$&$\kappa_3$&$\kappa_4$&$k_{eq}$&
$\frac{m_L^*}{M}$&$K$&$B$\\ \hline
S3&5.30&15.2&0.28&0.31& & & & &258.2&0.675&259&16.1\\ \hline
B1&5.7&15.3&0.28&0.30& & &0.5&-4.9&256.0&0.666&209&16.2 \\ \hline
B2&5.7&15.3&0.28&0.30&-0.055&0.18& & &257.3&0.661&201&16.1 \\ \hline
B3&5.6&15.27&0.28&0.30& & &0.31 &-4.1&259.1&0.659&185&16.1 \\ \hline 
B4&5.6&15.3 &0.28&0.31& & &0.9&-8.1  &256.4&0.669&191&16.1 \\ \hline
C1&5.7&15.3&0.28&0.30&-0.05&0.155& & &256.3&0.665&218&16.2 \\ \hline
C2&5.8&15.3&0.28&0.30&-0.11&0.35& & &256.1&0.662&161&16.2 \\ \hline
\end{tabular}
}
\end{table}

It is quite encouraging that the simple minimal model (\ref{toyBRPP}) 
with BR scaling captures
so much of the physics of nuclear matter. Of course, by
itself, there is no big deal in what is obtained by the truncated model: 
It is not a prediction. What is not trivial, however, is
that once we have a Lagrangian of the form (\ref{toyBRPP}) which defines the
mean fields, then we are able to control with some confidence the background 
around which we can fluctuate, which was the principal objective we set 
at the beginning of our paper. The power of the simple Lagrangian 
is that we can now 
treat fluctuations at {\it higher densities} 
as one encounters in heavy-ion collisions,  not just
at an equilibrium point. The description of such fluctuations does not
suffer from the sensitivity with which the EOS depends at $\rho >\rho_0$ 
on the many-body
correlation terms (\ref{nbody}). Some of these issues are illustrated 
in the next section.

\subsection{Some consequences}
\subsubsection{\it The $\omega$ in medium}
\indent\indent
Suppose one probes the propagation of an $\omega$ meson in nuclear medium
as in HADES or CEBAF experiments, say through dilepton production. The
$\omega$'s will decay primarily outside of the nuclear medium, but let us
suppose that experimental conditions are chosen so that the leptons
from the $\omega$ decaying inside dense matter
can be detected. See \cite{frimansoyeur}
for discussions on this issue. The question is whether the
dileptons will probe the BR-scaled mass or the quantity given by 
(\ref{omegaeff}). 
The behavior of the $\omega$ mass would differ drastically
in the two scenarios. 
A straightforward application of FTS1 theory would suggest that
at a density $\rho\lsim \rho_0$,
the $\omega$ mass as ``seen'' by the dileptons will increase slightly
instead of decrease. 
Since in FTS1 theory, the vector coupling $g_v$ does not scale, this means 
that $(g_v^\star/m_v^\star)$ will effectively decrease.
On the other hand if the vector 
coupling constant drops together with the mass
at increasing density as in the BR-scaling 
model\footnote{It is interesting that the dropping $\omega$ mass is also 
found in a recent QCD sum-rule calculation based on current
correlation functions by Klingl, Kaiser and Weise 
\cite{KKW} who, however, do not see the dropping of the $\rho$ mass.}, 
the situation could be quite
different, particularly if dileptons are produced at a density 
$\rho\sim 3\rho_0$ as
in the CERES experiments:
The $\omega$ will then be expected
to BR-scale up to the phase transition\footnote{It has been recently suggested
\cite{kurt} that at some high density, Lorentz symmetry can be spontaneously
broken giving rise to light $\omega$ mesons as ``almost Goldstone"
bosons. Such mesons could be a source of copious dileptons at some density
higher than normal matter density.}. 
Thus measuring the $\omega$ mass 
shift could be a key test of the BR scaling idea as opposed to the FTS1-type
interpretations. This interesting issue
will be studied in forthcoming experiments at GSI and CEBAF. 
\subsubsection{\it Nuclear static properties}
\indent\indent
Given the link between BR-scaled chiral Lagrangians and Fermi-liquid
fixed point theory, one should be able to make
a connection between the parameters that enter into such nuclear 
static properties as $\delta g_l$, referred to in the literature as 
the ``exchange-current" contribution to
the orbital gyromagnetic ratio, and the effective mass of the vector 
mesons $\omega$ and $\rho$. 
In this subsection, we shall show that this is indeed
possible. The results were already reported in \cite{FR96} but
we shall discuss them in the context developed in this paper.
The key element that is intimately
related to the Landau parameter $F_1$ is the universal scaling factor
$\Phi$, not the FTS1
effective mass discussed above that includes 
many-body correlations. To clarify this point, consider the Landau
effective mass of the nucleon $m^\star_L$ given in terms of the Landau
parameter $F_1$:
\be
\label{mstar}
\frac{m^\star_L}{m_N} = 1 + \frac{F_1}{3} = (1-\frac{\tilde{F_1}}{3})^{-1},
\label{landauM}\ee
where $ \tilde{F_1} = (m_N/m^\star_L) F_1$. Including the pion contribution,
we have a short-ranged term and a long-ranged term
\be
\tilde{F_1} = \tilde{F_1^\omega} +
\tilde{F_1^\pi}\label{F1}
\ee
where
\be
\label{fomega}
\tilde{F_1^\omega}&=&\frac{m_N}{m_L^\star} F_1^\omega=
-C_\omega^2\frac{2k_F^3}{\pi^2 m_N^\sigma },\\
\tilde{F_1^\pi}&=& -3\frac{m_N}{k_F}\frac{\der}{\der p}\Sigma_\pi (p)|_{p=k_F},
\ee
where the superscript denotes the relevant meson exchanged, $\Sigma_\pi$
is the nucleon self-energy (Fock term) involving one-pion exchange --
a non-local four-Fermi interaction -- and
\be
m_N^\sigma:=m_N\Phi
\ee
the BR-scaled nucleon mass in the absence of pions\footnote{Note that
$m_N^\sigma$ corresponds to $M^\star$ in the toy model with
BR scaling (\ref{toyBRPP}).}. 
It follows from the quasiparticle velocity at the Fermi surface \cite{FR96}
\be
\frac{d}{dp}\epsilon (p)|_{p=k_F}=\frac{k_F}{m_L^\star}=\frac{k_F}{m_N^\sigma}
+\frac{d}{dp}\Sigma_\pi (p)|_{p=k_F}
\ee
given by the BR-scaled Lagrangian together with 
Eqs. (\ref{landauM}) and (\ref{F1}) that
the $\omega$ contribution to the Landau
parameter $F_1$ is governed only by the factor $\Phi$:
\be
\tilde{F}_1^\omega
=3(1-\Phi^{-1}).\label{Phidefined}
\ee
This is the key relation that links the nucleon scaling present in mean field
theories to the scaling of the vector mesons in medium derived via chiral
symmetry plus scale anomaly. It is also this 
relation that connects the behavior of hadrons in heavy-ion collisions
to low-energy nuclear spectroscopic properties as we shall describe
below. Understanding of this relation would be crucial if one wanted to
have a unified description based on an effective chiral Lagrangian.

The scaling factor $\Phi (\rho)$ is known from the QCD sum rule calculation
for the in-medium mass of the $\rho$ meson at $\rho=\rho_0$ \cite{jinrho}
\be
\Phi (\rho_0)=0.78\pm 0.08
\ee
which can also be extracted from an 
in-medium Gell-Mann-Oakes-Renner formula for
the pion mass \cite{BRPR96}.
Since the contribution from the pion exchange is fixed by chiral
symmetry for a given density, i.e. at $\rho=\rho_0$,
\be
\frac 13
\tilde{F_1^\pi}&=& -\frac{3f_{\pi NN}^2m_N}{8\pi^2k_F}[\frac{m_\pi^2+2k_F^2}
{2k_F^2} \ln\frac{m_\pi^2+4k_F^2}{m_\pi^2}-2]\nonumber\\
&\approx& -0.153,\label{FockM}
\ee
the Landau mass for the nucleon
is entirely given once we assume the $\omega$ mass scales
\`a la BR\cite{BR96}:
\be
\frac{m^\star_L}{m_N}&=&\Phi\left(1 + \frac 13 F^\pi_1\right) \nonumber\\
&=& \left(\Phi^{-1}-\frac 13 \tilde{F}_1^\pi\right)^{-1}\nonumber\\
&=& (1/0.78 +0.153)^{-1}=0.69(7)\label{L}
\ee
which should be identified with the nucleon effective mass determined
by QCD sum rule at $\rho=\rho_0$\cite{jin}
\be
\frac{m_N^\star}{m_N}=0.69\pm {}^{0.06}_{0.14}.
\ee
The effective mass for the nucleon found with the toy model with BR scaling
(\ref{toyBRPP}) (with the set S3)
denoted there as $m_L^\star$, $m_L^\star/m_N \approx 0.68$,
is consistent with this QCD sum rule value. This provides 
one more support for our assertion.

The strongest support for this identification comes from the role that
the $\Phi$ factor plays in $\delta g_l$, the exchange-current correction
to the orbital gyromagnetic ratio of nuclei.
The response to a slowly-varying electromagnetic field of an odd
nucleon with momentum $\vec{p}$ added to a closed Fermi sea
can, in Landau theory, be represented by the current 
\be
\vec{J}= \frac{\vec{p}}{m_N}\left(\frac{1+\tau_3}{2} +\frac 16
\frac{F_1^\prime -F_1}{1+F_1/3} \tau_3\right) \label{qpcurrent}
\ee
where $m_N$ is the nucleon mass in medium-free space.
The long-wavelength limit of the current is not unique. 
The physically relevant one
corresponds to the limit $q \rightarrow 0, \omega \rightarrow 0$ with
$q/\omega \rightarrow 0$, where $(\omega,q)$ is the four-momentum transfer.
The current (\ref{qpcurrent}) defines the gyromagnetic ratio
\be
g_l=\frac{1+\tau_3}{2} +\delta g_l
\ee
where
\be
\delta g_l=\frac 16 \frac{F_1^\prime -F_1}{1+F_1/3}\tau_3
=\frac 16 (\tilde{F}_1^\prime-\tilde{F}_1)\tau_3.\label{deltalandau}
\ee
This expression is recovered simply if one calculates the exchange
of an $\omega$ and a $\rho$  with  BR scaling masses. 
The result  obtained recently in \cite{FR96} is 
\be
\delta g_l=\frac 49\left[\Phi^{-1} -1 -\frac 12 \tilde{F}_1^\pi\right]\tau_3.
\ee
This result is highly nontrivial in that (1) the $\omega$ contribution
restores the single-particle moment defined in terms of the free-space mass
$m_N$, not of the BR scaled mass\footnote{This is reminiscent of
the Kohn theorem for the cyclotron frequency of an electron in the metal
in a 
magnetic field where the free-space mass of the electron, not the
Landau mass, enters in the formula for the frequency.}, as required by
Ward identities and (2) the correction occurs {\it only}
in the 
isovector part. The numerical value for $\delta g_l$ at nuclear matter density
\be
\delta g_l=0.227\tau_3
\ee
agrees perfectly with the experimental value obtained from giant dipole
resonances in heavy nuclei\cite{schumacher}
\be
\delta g_l^p=0.23\pm 0.03.
\ee

We should emphasize that the link between the Landau parameter that
figures in the Fermi-liquid structure of nuclear matter and the BR scaling
that figures in an effective chiral Lagrangian supplies a stringent
consistency check of the theory. Another non-trivial
consistency check is given in the
strange-flavor sector which will be described below although the results have
been reported elsewhere.

\subsubsection{\it Fluctuations in the strange flavor direction}
\indent\indent
In considering kaonic fluctuations inside nuclear medium, the general argument
developed above suggests that we are to
take the ${\O} (Q^2)$ SU(3) chiral Lagrangian with BR-scaled parameters
and with bilinears in the baryon field taken 
in mean field. In the kaon direction, this then gives (modulo the 
``range term'' discussed below) for symmetric nuclear matter
\be
\L^{eff}_{KN}=\frac{-6i}{8\f^2}\overline{K}\del_t K\la {B}^\dagger B\ra +
\frac{\Sigma_{KN}}{\f^2} \overline{K}K \la \overline{B}B\ra
\equiv {\cal L}_\omega
+{\cal L}_\sigma\label{kaonL}
\ee
where $K^T=(K^+ K^0)$. The constant $\f$ in (\ref{kaonL}) can be identified
as the pion decay constant scaling as the square-root of the quark condensate
$\la\bar{q}q\ra$ \cite{BRPR96,BR91}. 
The appearance of $f_\pi^\star$ indicates the
BR-scaling.\footnote{As noted in \cite{BR96}, there can be
no non-derivative direct coupling between a Goldstone boson and a baryon 
like the second term of Eq. (\ref{kaonL}) in the chiral limit. 
Thus the direct coupling arises entirely through a chiral symmetry breaking
-- or quark masses -- in QCD. Pions couple nonderivatively to baryons
in the same way with the coefficient $\frac{\Sigma_{\pi N}}{{f_\pi^\star}^2}$.
In \cite{BR96}, this relation is given an interpretation in terms of the
$S$ exchange -- the identification exploited below.}
The potential felt by the kaon in the background
of nuclear matter is then given by 
\be
V_{K^\pm}&=&\pm\frac{3}{8{f^\star_\pi}^2}\rho,\label{vectorpot}\\
S_{K^\pm}&=&-\frac{\Sigma_{KN}}{2m_K \f^2}\rho_s\label{scalarpot}
\ee
where $\rho=\la B^\dagger B\ra$ and $\rho_s=\la\overline{B}B\ra$.
At nuclear matter density, $\rho=\rho_0$, we can identify these 
results as one third of the corresponding potentials for
nucleons, so we write
\be
V_{K^\pm}\approx\pm\frac 13 V_N\label{omegascale}
\ee
and
\be
S_{K^\pm}\approx \frac 13 S_N.
\ee
One way of understanding this result is that when written in terms of
BR scaling, we are essentially getting a {\it quasi-quark} description and
the factor 1/3 represents that the kaon carries 1/3 of the number of 
chiral quarks lodged in the nucleon. 
We expect the {\it quasi-quark} description to be good, once the meson mass
has decreased substantially with density as in the $K^-$ -case \cite{BR96}, 
but possibly not in the $K^+$ case where the mass does not move down with 
density. In the latter case the pseudo-Goldstone description should 
continue to be 
correct. (In particular, the range term is important for the $K^+$.)

Given Walecka-type mean fields for the 
nucleons, we can now calculate the corresponding
mean-field potential for $K^-$-nuclear interactions in symmetric nuclear
matter.
{}From the results obtained above, we have
\be
S_{K^-} +V_{K^-}\approx \frac 13 (S_N-V_N).
\ee
Phenomenology in Walecka-type mean-field theory gives
$(S_N-V_N)\lsim -600\ {\mbox{MeV}}$ for $\rho=\rho_0$ \cite{walecka}.
This leads to
the prediction that at nuclear matter density
\be
S_{K^-}+V_{K^-}\lsim -200\ {\mbox{MeV}}.
\ee
This seems to be consistent with the result of the analysis in K-mesic atoms
made by  Friedman, Gal and Batty
\cite{friedman} who find attraction at $\rho\approx 0.97\rho_0$ of
\be
S_{K^-}+V_{K^-}=-200\pm 20\ {\mbox{MeV}}.
\ee

As noted in \cite{BR96}, there is a correction called ``range term"
that appears at the same order of the chiral counting as the scalar potential (\ref{scalarpot}) which is proportional to second derivative
on the kaon field and hence $\sim \omega_K^2$ for an S-wave
kaon where $\omega_K$ is the
frequency of the kaon field. This correction can be approximately 
implemented by multiplying the scalar term in (\ref{scalarpot}) by
the factor $(1-0.37 \omega_K^2/m_K^2)$. With this correction, we find 
for $\rho=\rho_0$\cite{BR96}
\be
S_{K^-}+V_{K^-}\sim -192\ {\mbox{MeV}}.
\ee
For the $K^-$, the ``range correction" is not numerically significant. However
the situation is different for $K^+$-nuclear interaction. In fact, including
the range correction makes the $K^+$ effective mass to {\it increase} 
with density in contrast
to the $K^-$: We find the $K^+$ potential at nuclear
matter density to be effectively repulsive by the amount
\be
S_{K^+}+V_{K^+}\sim 25\ {\mbox{MeV}}\ \ \ {\mbox{at}} \ \ \rho=\rho_0.
\ee
\subsubsection{\it Going to higher densities in strange matter}
\indent\indent
The simple description given by the Lagrangian (\ref{kaonL}), corresponding
to the tree order with the BR-scaled Lagrangian, seems to work fairly
well up to $\rho\sim \rho_0$, but it must require corrections as density
is further increased. This is already indicated
in the construction of the BR-scaled chiral Lagrangian that reproduces 
FTS1 theory, i.e.,
the effective scaling of the coupling constant $g_v$ needed in describing
nuclear matter. More significantly, the Lagrangian (\ref{kaonL}), when
naively extrapolated to $\rho\sim 3\rho_0$, would be inconsistent with
what was observed in the KaoS kaon flow data \cite{gebgsi}.

The most efficient way to go higher in density is to bring in massive fields
in (\ref{kaonL}). To do this, one can think of the first term of (\ref{kaonL})
as arising from an $\omega$ exchange (and a $\rho$ exchange for nonsymmetric
nuclear matter) and the second term as coming from a scalar $\phi$ exchange.
This means that $1/{f_\pi^\star}^2$ in the first term is to be replaced,
in the notation of the Lagrangian (\ref{toyBRPP}),
by $\frac{{2}{g_v^\star}^2}{{m_v^\star}^2}$ and 
$\frac{\Sigma_{KN}}{{f_\pi^\star}^2}$ in the second term by 
$\frac{2m_K {h}^2}{3{m_s^\star}^2}$. (This also means that
the KSRF relation for the vector meson mass
cannot be naively applied in medium.)
From the foregoing discussion,
we expect that the first term will remain unscaled and the second term
scaled\footnote{In order to compare with the analysis of \cite{gebgsi},
one should note that $h$ is smaller (by about $1/2$) than the
scalar coupling in FTS1 theory. In addition, one should not forget
the ``range term" which tends to compensate the $1/\Phi^2$ scaling.}
as ${\Phi}^{-2}$ as one increases the density up to
the regime probed in the KaoS and FOPI experiments \cite{gebgsi}.
\subsubsection{\it Kaon condensation in compact-star matter}
\indent\indent
The Lagrangian (\ref{kaonL}) or more precisely the vector-meson-implemented
version of it was used in \cite{BR96} to calculate
the critical density for condensing $K^-$'s in dense neutron matter.
For this, nuclear matter informations provided by the FTS1 Lagrangian 
need to be  
supplemented by isovector degrees of freedom to describe the neutron matter
initially present in compact stars. This has not been yet worked out in 
terms of the FTS1 Lagrangian although this could be effectuated by 
incorporating the isovector vector mesons $\rho$ and $a_1$ into the 
FTS1 Lagrangian. Using the symmetry energy fitted at nuclear matter
density of heavy nuclei and extrapolating it to densities greater than the 
normal matter density\footnote{A recent realistic calculation of the
symmetry energy in the formalism of Dirac-Brueckner approach \cite{LKLB}
confirms the extrapolation (to a density $\rho\sim 3\rho_0$)
used in \cite{BR96,LBMR}.}, together with an estimate of the chemical potential
for the electron, the Lagrangian (\ref{kaonL}) predicts \cite{BR96,LLB97}
\be
\rho_c\lsim 3\rho_0.
\ee

That the critical density is of the order of a few times normal matter
density assures that the Lagrangian (\ref{kaonL}) is an appropriate one
since the same Lagrangian is checked in nuclear matter through
the heavy-ion experiments FOPI and KaoS up to $\rho\sim 3\rho_0$. 

But are there any important corrections missed in this treatment? 

To answer this question, we should note that
the mean field prediction made above contains certain
non-perturbative contributions that are not accessible in low-order 
chiral perturbation expansion. For instance, in \cite{LBMR} where
the critical density is calculated to order $Q^3$ (or one-loop order)
in chiral perturbation theory, one out
of two constants that appear in the four-Fermi interaction terms in the 
Lagrangian was fixed to reproduce the Friedman-Gal-Batty attraction 
of 200 MeV in the 
kaonic atom data as does the Lagrangian (\ref{kaonL}). 
Thus it invokes an ingredient that is not directly extracted
from the set of available on-shell data. Indeed a recent calculation 
\cite{WRW} to
${{\cal O} (Q^2)}$ in chiral perturbation theory that is highly constrained by
the ensemble of on-shell kaon-nucleon data and that includes both Pauli and 
short-range correlations for many-body effects is found to give at most 
about 120 MeV attraction at nuclear matter density. Thus the crucial input
here is the strength of the $K^-$-nuclear interaction in dense medium.
If the analysis of the $K$-mesic atom by Friedman et al indicating
the 200 MeV attraction turned out to be incorrect and the attraction
 came down to
$100\sim 120$ MeV as found in \cite{WRW}, this would give a strong constraint
on the constants that enter in four-Fermi interactions in the chiral
Lagrangian. This would presumably account for the need for a dropping 
vector coupling $g_v^\star$ required for $\rho\gsim \rho_0$.
This crucial information is also expected to come from on-going heavy-ion
experiments.
\section{Summary and Conclusions}
\indent\indent
In this paper, a first attempt is made to go from an effective
(quantal) chiral Lagrangian to an effective field theory for nuclear matter
at variable densities, with the aim to build a bridge between (low-energy)
nuclear spectroscopic properties under normal condition and (higher-energy) 
physics of dense matter under extreme conditions expected to be found in 
relativistic heavy-ion collisions and in compact stars such as neutron stars. 
A construction of this sort
will be necessary to eventually understand the QCD phase transition(s) believed 
to take place at high temperature and/or high density.
For this purpose, we take
the FTS1 (the effective chiral Lagrangian model of Furnstahl et al) 
\cite{tang} which is found to be highly successful in the phenomenology
of finite nuclei and nuclear matter, to argue that an effective
chiral Lagrangian constructed in high chiral orders corresponds, {\it 
in mean field}, to Lynn's chiral soliton \cite{lynn} with chiral liquid 
structure. This provides an efficient background around which quantum 
fluctuations can be reliably calculated.

Next, using the renormalization group-flow arguments developed in condensed
matter physics, we proceed to propose that the chiral liquid theory 
with the FTS1 Lagrangian (in mean field) corresponds to Landau's
Fermi liquid fixed point theory \cite{shankar,froehlich}. We develop the notion that the FTS1 
theory in mean field is at fixed points except for the scalar sector which develops a large anomalous dimension which we attribute to
a strong coupling situation. We then suggest that the 
strong-coupling theory with the parameters defined in matter-free space
can be transformed into a weak-coupling theory if the chiral Lagrangian 
is rewritten in terms of BR-scaled parameters. We construct a simple model with BR-scaled masses 
that gives a fairly good description of ground-state properties 
with fits comparable to the full FTS1 theory. The simple BR-scaling Lagrangian provides the background  
at an arbitrary density 
around which
fluctuations can be calculated with the
tree diagrams yielding the dominant contributions. We have thus 
obtained a quasiparticle picture of a strongly correlated system at
densities away from the equilibrium point.

The identification of the BR-scaling parameter $\Phi$ with the Landau-Migdal
Fermi liquid parameter $F_1$ leads to a set of relations
that connect the physics that governs heavy-ion collisions, e.g.,
the CERES dilepton data and the nucleon and kaon flow data of FOPI and 
KaoS etc.
to such low-energy spectroscopic properties as effective (Landau)
nucleon mass, effective $g_A$ and {the exchange-current correction to the} 
orbital gyromagnetic ratio, $\delta
g_l$ etc. These relations are found to be satisfied to a surprising
accuracy. Finally the formalism allows a consistent calculation of
kaon condensation in dense star matter which is proposed to play an 
important role in supernovae explosion with the remnant forming
``nucleon" or ``nuclear" 
stars or going into small black holes \cite{brownbethe,LLB97}.

While for an exploration, our results are satisfying, 
there are several crucial links
that remain conjectural in the work and require a lot more work. 
We have not yet established in
a convincing way that a nontopological soliton coming from a high-order
effective chiral Lagrangian accurately describes
 nuclear matter that we know of.
The first obstacle here is that a realistic effective Lagrangian that
contains sufficiently high-order loop corrections including non-analytic
terms has not yet been constructed. Lynn's argument for the existence of
such a soliton solution and identification with a drop of nuclear matter
is based on a highly truncated Lagrangian (ignoring non-analytic terms).
We are simply assuming that the FTS1 Lagrangian is a sufficiently realistic
version (in terms of explicit vector and scalar degrees of freedom
that are integrated out by Lynn) 
of Lynn's effective Lagrangian. To prove that this assumption is valid
is an open
problem. Our argument for interpreting the FTS1 with the anomalous
dimension $d_{an}\approx 5/3$ for the quarkonium scalar field as a 
strong-coupling theory which can be reinterpreted in terms of 
a weak-coupling theory expressed with BR scaling is heuristic at best
and needs to be
sharpened, although our results strongly indicate that it is correct.
Furthermore transcribing
the renormalization-group arguments developed in condensed matter physics
to dense hadronic matter -- involving more degrees of freedom and
more length scales --
remains to be made rigorous. This is an issue which is of the same nature
as transcribing Landau Fermi liquid theory to nuclear matter as in the work 
of Migdal and also as going from relativistic mean field
theory of Walecka type to Landau Fermi liquid theory as in the work of Matsui 
and others. 

Finally, there is the practical question as to 
how far in density the predictive power of the
BR-scaled effective Lagrangian can be pushed. In our simple numerical
calculation, we used a parameterization for the scaling function $\Phi (\rho)$
of the simple geometric form which can be valid, if at all, up to the normal 
matter density as seems to be supported by QCD sum-rule and
dynamical model calculations. At higher densities, the form used has no reason
to be accurate. By using the empirical information coming from nucleon and kaon
flows, one could infer its structure up to, say,
 $\rho\sim 3 \rho_0$ and if our
argument for kaon condensation is correct -- and hence kaon condensation
takes place at $\rho\lsim 3\rho_0$, then this will be good enough to
make a prediction for the critical density for kaon condensation. In 
calculating compact-star properties in supernovae explosions, however, the EOS
for densities considerably higher than the normal matter density, say, $\rho
\gsim 5\rho_0$ is required. 
It is unlikely that this high density can be accessed 
within the presently employed approximations. Not only will the structure
of the scaling function $\Phi$ be more complicated
but also the correlation
terms that are small perturbations at normal density may no longer be
so at higher densities, as pointed out by Pandharipande et al.
\cite{panda} who approach the effect of correlations from the high-density
limit. In particular, the notion of the scaling function $\Phi$ will have to
be modified in such a way that it will become a non-linear function of the
fields that figure in the process. This would alter the structure of the
Lagrangian field theory.
Furthermore there may be a phase transition (such as spontaneously broken
Lorentz symmetry, Georgi
vector limit, chiral
phase transition or meson condensation) lurking nearby in which case the 
present theory would have already broken down. 
These caveats will have to be carefully
examined before one can extrapolate the notion of BR scaling to
a high-density regime as required
for a reliable calculation of the compact-star
structure.

\subsection*{Acknowledgments}
\indent\indent
We are grateful to Bengt Friman for helpful correspondence on some
of the matters treated in this paper and to Chang-Hwan Lee for comments and
suggestions.
Parts of this paper were written while one of the authors (MR) was a
Humboldtpreistr\"ager at the Institute of Theoretical Physics of
the Technische Universit\"at M\"unchen (Munich, Germany) and an  
 ``IBERDROLA  de Ciencia y Tecnologia Professor" at the Department of
Theoretical Physics of the Universitat de Val\`encia (Valencia, Spain).
He is grateful to the two foundations for the financial support
and to Wolfram Weise (Munich) and Vicente Vento (Valencia)
for the warm hospitality. The work of GEB is supported in part
by the U.S. Department of Energy under the grant No. DE-FG02-88ER40388
and the work of CS and DPM 
by KOSEF through the Center for Theoretical Physics, Seoul National 
University, and in part by Korea Ministry of Education (BSRI97-2418).
\vskip 1cm
\pagebreak
\section*{Appendix: Effect of Many-Body Correlations on EOS}
\indent\indent
In this appendix, we compare various parameter sets with the FTS1, with a focus
on the parameters B2 and B4 that lead to instability 
in the system at
a slightly higher density than nuclear matter. Note that the equilibrium 
properties are well described by both the parameters that 
give stable EOS and those that do not at a density slightly above $\rho_0$.

\begin{figure}[tbh]
\setlength{\epsfysize}{6.in}
\centerline{\epsffile{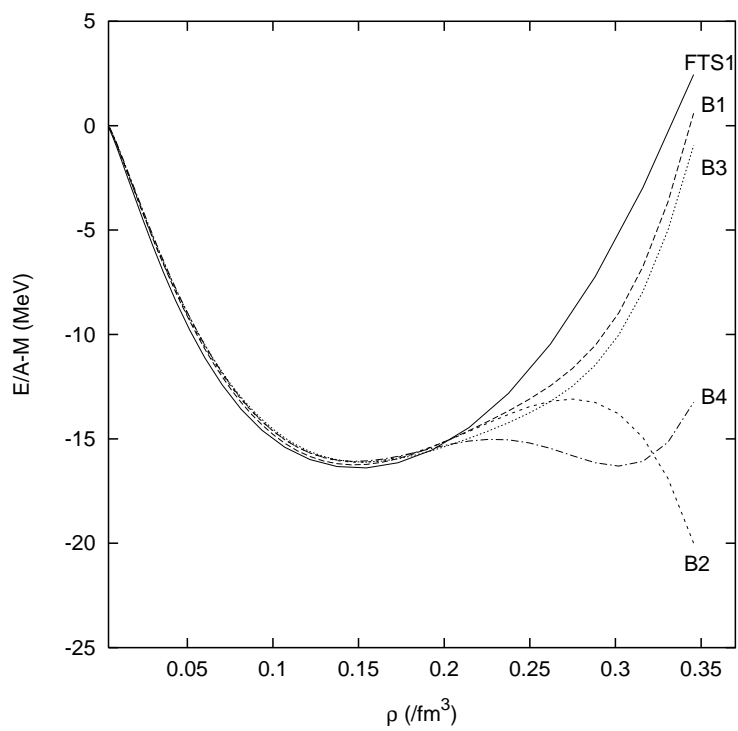}}
\caption[com]{$E/A-M$ vs. $\rho$ for the B1, B2, B3 and B4 models given in 
Table \ref{fit} compared with FTS1 theory. }
\label{comparison}
\end{figure}

\end{document}